\documentstyle[prl,aps,preprint,tighten]{revtex}
\begin{document}
\draft
\preprint{\parbox{2.2in}{FERMILAB--Pub--93/255--T\\hep-ph/9308337\\[12pt]}}
\title{Properties of Orbitally Excited Heavy-Light Mesons}
\author{Estia J. Eichten,
\thanks{Electronic address: (internet) eichten@fnal.fnal.gov}
Christopher T. Hill,
\thanks{Electronic address: (internet) hill@fnal.fnal.gov}
and Chris Quigg
\thanks{Electronic address: (internet) quigg@fnal.fnal.gov}
}
\address{Fermi National Accelerator Laboratory,
Batavia, Illinois 60510}
\date{\today}
\maketitle
\begin{abstract}
Orbitally excited heavy-light mesons are potentially important as tools for
tagging the flavors and momenta of ground-state pseudoscalars detected
through weak decays. We use heavy-quark symmetry supplemented by insights
gleaned from potential models to estimate masses and widths of $p$-wave
$B$, $B_s$, and $D_s$ mesons. We generalize these results to higher
excitations.
\end{abstract}

\narrowtext
Incisive study of particle-antiparticle mixing and $CP$-violation for
neutral $B$ mesons requires that the quantum numbers of the meson be
identified at the time of production.  That identification can be made by
observing the decay of a $B$ or $\bar{B}$ produced in association with a
particle of opposite $b$-number whose decay signals the flavor of the
neutral $B$ of interest.  The efficiency of flavor identification might
be considerably enhanced if the neutral $B$ under study were self-tagging
\cite{tags}.

Charmed mesons have been observed as (strong) decay products of orbitally
excited $(c\bar{q})$ states, through the decays $D^{**}\rightarrow \pi D$
and $D^{**}\rightarrow \pi D^*$ \cite{charm}.  The charge of the pion
emitted in the strong decay signals the flavor content of the charmed
meson.  If significant numbers of $B$ mesons are produced through
one or more narrow excited $(\bar{b}q)$ states, the strong decay
$B^{**\pm} \rightarrow B^{(*)0}\pi^\pm$
tags the neutral meson as $(\bar{b}d)$ or $(b\bar{d})$, respectively.

The primary application of $B^{**}$-tagging would be in the search for the
expected large $CP$-violating asymmetry in $(B^0~{\rm or}~\bar{B}^0)\!
\rightarrow\!J/\psi K_S$ decay \cite{isi}.
$B^{**}$-tagging may also resolve kinematical ambiguities in semileptonic
decays of charged and neutral $B$ mesons by choosing between two solutions
for the momentum of the undetected neutrino. In hadron colliders and
$Z^0$-factories,
kinematic tagging may make practical a high-statistics determination of
the form factors in semileptonic weak decay, and might ultimately allow
precise measurements of $V_{cb}$ and $V_{ub}$
\cite{Mats,CTH}.  The study of $B_s$-$\bar{B}_s$
mixing would be made easier if the kaon charge in the decay
$B^{***}\rightarrow K^\pm (B_s~{\rm or}~\bar{B}_s)$ served as a flavor tag.
Overall, efficient $B^{**}$-tagging would dramatically enhance the
prospects for studying $CP$-violation and $B_s$-$\bar{B}_s$ mixing.

In this Letter, we estimate the masses, widths, and branching fractions of
orbitally excited $B$, $D_s$, and $B_s$ states from the properties of
corresponding $K$ and $D$ levels. Our results show that one requirement for
the utility of $B^{**}$-tagging, narrow resonances, is likely to be met by
the $B_2^*$ and $B_1$. Experiment must rule on the strength of these lines
and the ratio of signal to background.

For hadrons containing a heavy quark $Q$, quantum chromodynamics displays
additional symmetries in the limit as the heavy-quark mass $m_Q$ becomes
large compared with a typical QCD scale \cite{hqet}. These heavy-quark
symmetries are powerful aids to understanding the spectrum and decays of
heavy-light $(Q\bar{q})$ mesons. Because $m_b \gg \Lambda_{\rm QCD}$,
heavy-quark symmetry should provide an excellent description of the $B$
and $B_s$ mesons. It is plausible that properties of $D$ mesons,
and even $K$ mesons, should also reflect approximate heavy-quark
symmetry.

One essential idea of the heavy-quark limit is that the
spin $\vec{s}_Q$ of the heavy quark and the total (spin + orbital) angular
momentum
$\vec{\jmath}_q = \vec{s}_q + \vec{L}$ of the light degrees of freedom are
separately conserved \cite{spin}. Accordingly, each energy level in the
excitation spectrum of $(Q\bar{q})$ mesons is composed of a degenerate pair
of states characterized by $j_q$ and the total spin
$\vec{J}=\vec{\jmath}_q+\vec{s}_Q$,
i.e., by $J=j_q\pm { \frac{1}{2}}$.  The ground-state
pseudoscalar and vector mesons, which are degenerate in the heavy-quark
limit, correspond to $j_q = { \frac{1}{2}}$, with $J=0$
and $1$.
Orbital excitations lead to two distinct doublets associated with
$j_q = L\pm{ \frac{1}{2}}$.

{\em Masses.\/}  The leading corrections to
the spectrum prescribed by heavy-quark symmetry are inversely
proportional to the heavy-quark mass.  We may write the mass of a
heavy-light meson as
\begin{equation}
        M(n{\rm L}_J(j_q))\! =\! M(1{\rm S}) + E(n{\rm L}(j_q)) +
        \frac{C(n{\rm L}_J(j_q))}{m_Q}\;,
        \label{mass}
\end{equation}
where $n$ is the principal quantum number and
$M({\rm 1S})=[3M(1{\rm S}_1)+M(1{\rm S}_0)]/4$ is the mass
of the ground state.  The excitation energy
$E(n{\rm L}(j_q))$ has a weak dependence on the heavy-quark mass.

Let us focus first upon the $j_q={ \frac{3}{2}}$ states
observed as narrow $D\pi$ or $D^*\pi$ resonances.  We will show below
that their counterparts in other heavy-light systems should also be narrow.
Our overall strategy is to use the observed properties of the $K$ and $D$
mesons to predict the properties of the orbitally excited $B$, $D_s$, and $B_s$
mesons. Experimental knowledge of the reference systems is still
fragmentary, so we must appeal to potential models to estimate how the
excitation spectrum varies with heavy-quark mass. Although nonrelativistic
potential models have obvious limitations for systems that include light
quarks, we find that the Buchm\"{u}ller-Tye potential \cite{BT} gives a
good account of the observed $K$, $D$, and $D_s$ levels. The
potential-model spectra can also serve as templates for unobserved states,
particularly those along the leading Regge trajectory.

According to Eq. (\ref{mass}), the masses of the strange and charmed
mesons with $j_q={ \frac{3}{2}}$ are given by
\begin{eqnarray}
        M(2{\rm P}_2)_K-M(1{\rm S})_K & = &
        E(2{\rm P})_K +\frac{C(2{\rm P}_2)}{m_s}\;\;\; ,
        \nonumber \\
        M(2{\rm P}_1)_K-M(1{\rm S})_K & = &
        E(2{\rm P})_K +\frac{C(2{\rm P}_1)}{m_s}\;\;\; ,
        \nonumber \\
        M(2{\rm P}_2)_D-M(1{\rm S})_D & = &
        E(2{\rm P})_D +\frac{C(2{\rm P}_2)}{m_c}\;\;\; ,
        \label{master} \\
        M(2{\rm P}_1)_D-M(1{\rm S})_D & = &
        E(2{\rm P})_D +\frac{C(2{\rm P}_1)}{m_c}
        \nonumber \;\;\; ,
\end{eqnarray}
where we have suppressed the $j_q$ label for brevity.
Upon identifying $E(2{\rm P})_D = E(2{\rm P})_K -\delta$,
where $\delta=32~{\rm MeV}$ is determined from the potential-model
spectra, we are left with four linear equations in the five unknowns
$E(2{\rm P})_K$, $C(2{\rm P}_2)$, $C(2{\rm P}_1)$,
        $m_s^{-1}$, and $m_c^{-1}$.

The $K$- and $D$-meson masses we use as experimental inputs are displayed
in Table \ref{masspredictions}. There is no ambiguity about the
$2^+(\frac{3}{2})$ levels. We identify $D_1(2424)$ as a $j_q=\frac{3}{2}$
level because it is narrow, as predicted \cite{IW,FL} by heavy-quark
symmetry. We follow Ito et al. \cite{K} in identifying $K_1(1270)$
as the $1^+(\frac{3}{2})$ level, because that assignment gives a consistent
picture of masses and widths.

To proceed, we choose a value for the charmed-quark mass, $m_c$. After
solving Eqs. (\ref{master}), we verify the reasonableness of $m_s$ and
predict the $j_q\!=\!{\frac{3}{2}}$ masses for the $B$, $D_s$, and $B_s$
families. We consider two sets of parameters inspired by $J/\psi$ and
$\Upsilon$ spectroscopy: $m_c\!=\!1.48~{\rm GeV}$, $m_b\!=\!4.8~{\rm GeV}$
\cite{BT}; and $m_c\!=\!1.84~{\rm GeV}$, $m_b\!=\!5.18~{\rm GeV}$
\cite{Cornell}. Both solutions
                [$C(2{\rm P}_2) \!=\! (0.0495, 0.06155)~{\rm GeV}^2$,
        $C(2{\rm P}_1) \!=\! (-0.0029, -0.00358)~{\rm  GeV}^2$,
        $E(2{\rm P})_K \!=\! (0.4844, 0.48445)~{\rm GeV}$,
        $m_s\!=\!(0.33, 0.41)~{\rm GeV}$]
yield reasonable values for the strange-quark mass.  Their implications
for the $B$, $D_s$, and $B_s$ levels are
consistent within $2~{\rm MeV}$.  The average values are presented in
Table \ref{masspredictions}.  Including the variation of excitation
energy represented by the parameter $\delta$ has lowered the
masses by 7, 26, and $32~{\rm MeV}$ for the $B$, $D_s$, and $B_s$ states.

Our prediction for the $1^+$ $D_s$ meson lies $34~{\rm MeV}$ below the
level observed \cite{PDG} at $2536.5 \pm 0.8~{\rm MeV}$. The spin-parity
assignment for this state as $1^+$, rather than $2^+$, is suggested by the
nonobservation of a $KD$ decay mode and supported by a recent analysis of
the helicity-angle distribution of the $D^{*0}$ in the decay
$D_{s1}\rightarrow D^{*0}K$ \cite{CLEO1+}. We take the
discrepancy between calculated and observed
masses as a measure of the limitations of our method.

The $2{\rm P}({ \frac{1}{2}})$ $D$ mesons have not yet been observed, so we
cannot predict the masses of other heavy-light states by this technique.
Splitting within the multiplet can be estimated using Eq. (\ref{mass})
from the kaon spectrum alone. The small splitting between
$K_0^*(1429)$ and $K_1(1402)$ implies that the $1^+({ \frac{1}{2}})$ and
$0^+({\frac{1}{2}})$ levels should be nearly degenerate in all the
heavy-light systems. Chiral symmetry and heavy-quark symmetry combined
suggest that, like their counterparts in the strange-meson spectrum, the
heavy-light $j_q\!=\!{ \frac{1}{2}}$ $p$-wave states should have large
widths for pionic decay to the ground states
\cite{chiralHQS}. This will make the discovery and study of these states
challenging, and will limit their utility for $B^{**}$-tagging.

{\em Decay widths.\/}  Consider the decay of an excited heavy-light
meson $H$, characterized by ${\rm L}_J(j_q)$, to a heavy-light meson
$H^\prime({\rm L}^\prime_{J^\prime}(j_q^\prime))$, and a
light hadron $h$ with spin $s_h$.  The amplitude for the emission of $h$
with orbital angular momentum $\ell$ relative to $H^\prime$ satisfies
certain symmetry relations because the decay dynamics become independent
of the heavy-quark spin in the $m_Q\!\rightarrow\!\infty$ limit of QCD
\cite{IW}.  The decay amplitude can be factored \cite{FL} into a reduced
amplitude ${\cal A}_{\rm R}$ times a normalized 6-$j$ symbol,
\begin{displaymath}
       {\cal A}(H\!\rightarrow\! H^\prime h) =
       (-1)^{s_Q+j_h\!+J^\prime\!+j_q\,}
       {\cal C}
       ^{s_Q,j_q^\prime,J^\prime}_{j_h,J,j_q}\!\!{\cal A}_{\rm
       R}(j_h,\ell,j_q,j_q^\prime) ,
\end{displaymath}
where $\!{\cal C}^{s_Q,j_q^\prime,J^\prime}_{j_h,J,j_q}\!\!\!=\!
\sqrt{(2J^\prime+1)(2j_q+1)} \left\{\!\begin{array}{@{}ccc@{}}
s_Q & j_q^\prime & J^\prime \\ j_h & J & j_q  \\ \end{array}\!\right\}$
and $\vec{\jmath}_h \equiv \vec{s}_h+\vec{\ell}$.
The coefficients ${\cal C}$ depend only upon the total angular momentum $j_h$
of the light hadron, and not separately on its spin $s_h$ and the orbital
angular momentum wave $\ell$ of the decay.
The two-body decay rate may be written as
\begin{equation}
        \Gamma_{j_h,\ell}^{H\rightarrow H^\prime h} =
        ({\cal C}^{s_Q,j_q^\prime,J^\prime}_{j_h,J,j_q})^2
        p^{2\ell+1} F_{j_h,\ell}^{j_q,j_q^\prime}(p^2) ,
        \label{widthmaster}
\end{equation}
where $p$ is the three-momentum of the decay products in the rest frame
of $H$.  Heavy-quark symmetry does not predict the reduced
amplitude ${\cal A}_{\rm R}$ or the related
$F_{j_h,\ell}^{j_q,j_q^\prime}(p^2)$ for a particular decay.  Once
determined from the charmed or strange mesons,
these dynamical quantities may be used to predict related decays,
including those of orbitally excited $B$ mesons.  For each independent
decay process, we assume a Gaussian form
\begin{equation}
        F_{j_h,\ell}^{j_q,j_q^\prime}(p^2) =
        F_{j_h,\ell}^{j_q,j_q^\prime}(0) \exp(-p^2/\kappa^2) ,
        \label{gauss}
\end{equation}
and determine the overall strength of the decay and the momentum scale of
the form factor by fitting to existing data.  Our ability to predict
decay rates depends on the quality of the information used to set these
parameters.

In writing (\ref{widthmaster}) we have ignored $1/m_Q$
corrections to heavy-quark symmetry predictions for decay rates, except
as they modify the momentum $p$ of the decay products.  We assume that
the momentum scale $\kappa$ of the form factor in (\ref{gauss}) is
typical of hadronic processes ($\approx 1~{\rm GeV}$) and that it varies
little with decay angular momentum $\ell$.

The decays $2{\rm P}({ \frac{3}{2}}) \rightarrow 1{\rm S}
({\frac{1}{2}})+\pi$ are governed by a single $\ell=2$ amplitude. To
evaluate the transition strength $F_{2,2}^{\frac{3}{2},\frac{1}{2}}(0)$, we
fix $\Gamma(D_2^*\rightarrow D\pi)+\Gamma(D_2^*\rightarrow D^*\pi) =
25~{\rm MeV}$, as suggested by recent experiments \cite{charm}. This
determines all the pionic transitions between the $2{\rm P}({
\frac{3}{2}})$ and $1{\rm S}({\frac{1}{2}})$ multiplets. The results are
shown in Table \ref{pwavewidths}, where we indicate the variation of the
predicted rates as the momentum scale $\kappa$ ranges from 0.8 to 1.2~GeV.
The strengths of $K$ and $\eta$ transitions are determined by $SU(3)$
\cite{etacomp}. The predictions agree well with what is known about the
$L=1$ $D$ and $D_s$ states. The ratio $\Gamma(D_2^*\rightarrow D\pi) /
\Gamma(D_2^*\rightarrow D^*\pi) = 1.8$ is consistent with the Particle Data
Group average, $2.4\pm0.7$ \cite{PDG}, and with a recent CLEO measurement,
$2.1 \pm 0.6 \pm 0.6$ \cite{CLEO93}.

Increasing the $D_{s1}$ and $D_{s2}^*$
masses by 34~MeV to match the observations of $D_{s1}$ increases each of
the partial widths for those states by 1 or 2~MeV.  The narrow width
observed for $D_{s1}$ is close to the prediction from heavy-quark
symmetry.  This suggests that mixing of the narrow
$2{\rm P}({\frac{3}{2}})$ level with the broader
$2{\rm P}({\frac{1}{2}})$ state \cite{IW,FL} is negligible. This pattern
should hold for $B$ and $B_s$ as well. We have also applied heavy-quark
dynamics to the decays of the $2{\rm P}({ \frac{3}{2}})$ strange mesons.
The pionic transition rates given in Table \ref{pwavewidths} are in
surprisingly good agreement with experiment.

Decays of the $2{\rm P}({\frac{3}{2}})$ states into a vector meson plus a
$1{\rm S}({\frac{1}{2}})$ level are governed by three independent decay
amplitudes characterized by $(j_h,\ell)=(2,2)$, $(1,2)$, and $(1,0)$.
$SU(6)$ symmetry identifies the $(2,2)$ transition strength with the
$F_{2,2}^{\frac{3}{2},\frac{1}{2}}(0)$ for pion emission. The two new
amplitudes occur in a fixed combination that should be dominated by the
$\ell=0$ amplitude. We have to evaluate one new transition strength,
$F_{1,0}^{\frac{3}{2},\frac{1}{2}}(0)$. Lacking measurements of partial
widths for vector meson emission in the charmed states, and encouraged by
the satisfactory pattern of pionic decay widths for the strange resonances,
we use the decay rate $\Gamma(K_1(1270)\rightarrow \rho+K)= 37.8~{\rm MeV}$
to fix $F_{1,0}^{\frac{3}{2},\frac{1}{2}}(0)$. We smear the expression
(\ref{widthmaster}) for the partial width over a Breit-Wigner form to take
account of the 150-MeV width of the $\rho$ resonance.

The resulting estimates for the $\rho$ transitions are also shown in Table
\ref{pwavewidths}. The dependence on $\kappa$ is much more pronounced than
for the pseudoscalar transitions because of the wide variation in momentum
over the $\rho$ peak. Rates for $K^{**}\!\rightarrow\! K\omega$ decays follow
from $SU(3)$ symmetry.

The results collected in Table \ref{pwavewidths} show that both the
$B_2^*$ and the $B_1$ states should be narrow (20 to 40~MeV), with large
branching fractions to a ground-state $B$ or $B^*$ plus a pion.  These
states should also have significant two-pion transitions that we have
modeled by the low-mass tail of the $\rho$ resonance.  The strange
states, $B_{s2}^*$ and $B_{s1}$, are very narrow
($\Gamma{\raisebox{-.4ex}{\rlap{$\sim$}} \raisebox{.4ex}{$<$}} 10~{\rm MeV}$);
their dominant decays are by kaon emission to the ground-state $B$ and
$B^*$.  The consistent picture of $K_1$ and $K_2^*$ decay rates supports
the identification \cite{K} of $K_1(1270)$ as the
$2{\rm P}_1({\frac{3}{2}})$ level.

To assess the prospects for tagging $B_s$, we consider briefly the $d$-wave
heavy-light mesons with $j_q\!=\!{\frac{5}{2}}$.  Only the $K$ mesons have been
observed. The identification of the $K_3^*(1770)$ as a
$3{\rm D}_3({\frac{5}{2}})$ level is clear.
Two $J^P = 2^-$ levels, $K_2(1773)$ and $K_2(1816)$, are candidates for
its partner \cite{LASS2-}. Whatever the assignment for the $3{\rm
D}_2({\frac{5}{2}})$ level, the splitting within the
$j_q\!=\!{\frac{5}{2}}$ doublet will be very small for the $D$,
$B$, $D_s$, and $B_s$ systems.  We use the Buchm\"{u}ller-Tye potential
\cite{BT} to estimate the masses of the $L=2$ heavy-light states shown in
Table \ref{shortdwavewidths}.

To evaluate the transition strength
$F_{3,3}^{\frac{5}{2},\frac{1}{2}}(0)$ for pseudoscalar emission, we fix
$\Gamma(K_3^*\rightarrow K^* \pi)=45~{\rm MeV}$.  As before,
$SU(6)$ symmetry determines the strength
$F_{3,3}^{\frac{5}{2},\frac{1}{2}}(0)$ for vector meson emission.  In the
absence of measurements that would allow us to fix the other important
decay amplitude, we have set
$F_{2,1}^{\frac{5}{2},\frac{1}{2}}(0)=0$.  Our projections for
vector-meson emission will therefore be underestimates.  We summarize our
expectations for the total widths of the $3{\rm D}({\frac{5}{2}})$
states in Table \ref{shortdwavewidths}.

The $3{\rm D}({ \frac{5}{2}})$ $B$ mesons will be broad (250 to
400~MeV), but decay with about thirty percent probability to $B_s$ and
$B_s^*$ by emission of a $K$ or $K^*$. The estimate for the branching
fraction is less sensitive than the widths to variations
in $\kappa$. The favorable branching fraction means that it
might be possible to use $B_3^*$ and $B_2$ decays to tag the $B_s$,
in spite of the very large total widths.

Properties of orbitally excited heavy-light mesons will test the
validity of heavy-quark symmetry, which may offer new insight into the
spectrum of strange mesons.  If the narrow $B_2^*$ and $B_1$ are
copiously produced with little background, efficient tagging of flavor
and momentum may be at hand.  Prospects for incisive $B$ studies at high
energies would then be dramatically enhanced \cite{CTH}.

We thank Joel Butler, Shekhar Shukla, Paris Sphicas, and especially Tom
LeCompte for stimulating discussions. This work was performed at the Fermi
National Accelerator Laboratory, which is operated by Universities Research
Association, Inc., under contract DE-AC02-76CHO3000 with the U.S.
Department of Energy.

\newpage
\mediumtext
\begin{table}
 \caption{Masses (in MeV) predicted for the 2P$({\frac{3}{2}})$ levels of
the $B$, $D_s$, and $B_s$ systems. Underlined entries are Particle Data
Group averages \protect \cite{PDG} used as inputs.}
        \begin{tabular}{lccccc}
                Meson Family & $K$ & $D$ & $B$ & $D_s$ & $B_s$  \\
                \hline
                $M(1{\rm S})$ & \underline{794.3} & \underline{1973.2} &
                \underline{5313.1} & \underline{2074.9} &
5409.1\tablenote{Assuming
                that $M(1{\rm S})=M(1{\rm S}_0) + 34.5~{\rm MeV}$, as in the
$B$
                system.  The pseudoscalar mass,
        $M_{B_s}=5374.6~{\rm MeV}$, is the weighted mean of the ALEPH and CDF
values
        \cite{B_s mass}.} \\
                Level Shift $\delta$ & 0 & 32 & 42 & 56 & 67  \\
                \hline
                $M(2^+({ \frac{3}{2}}))$ &
                \underline{$1429 \pm 6$} & \underline{$2459.4 \pm 2.2$} & 5767
                & 2537 & 5846  \\
                $M(1^+({ \frac{3}{2}}))$ &
                \underline{$1270 \pm 10$} & \underline{$2424 \pm 6$} & 5755
                & 2502 & 5834  \\
                $M(2^+({ \frac{3}{2}}))-M(1^+({ \frac{3}{2}}))$ &
                 159 & 35 & 12 & 35 & 12  \\
        \end{tabular}
        \protect\label{masspredictions}
\end{table}
\narrowtext
\begin{table}
        \caption{Decay rates of the
        2P$({\frac{3}{2}})$ heavy-light mesons.}
        \begin{tabular}{lcc}
         & \multicolumn{2}{c}{Width (MeV)} \\
Transition & Calculated & Observed\tablenote{1992 Particle Data Group values
 \protect\cite{PDG}.}\\
\hline
$D_2^*(2459) \rightarrow D^*\pi$    & 9\tablenote{Sum fixed at 25 MeV.} & \\
$D_2^*(2459) \rightarrow D\pi$  & $16^{\rm b}$ & \\
$D_2^*(2459) \rightarrow D\eta$  & $\sim 0.1$ & \\
$D_2^*(2459) \rightarrow D \rho$    & 5 to 13 & \\
\hline
$D_2^*(2459) \rightarrow {\rm all}$ &  30 to 38 & $19 \pm 7$ \\[2pt]
$D_1(2424) \rightarrow D^*\pi$  & 11 to 13 & \\
$D_1(2424) \rightarrow D \rho$  & 8 to 11 & \\
\hline
$D_1(2424) \rightarrow {\rm all}$   &  19 to 23 & $20^{+9}_{-5}$ \\[4pt]
$D_{s2}^*(2537) \rightarrow D^*K$   & 2 to 4 & \\
$D_{s2}^*(2537) \rightarrow DK$ & 6 to 7 & \\
$D_{s2}^*(2537) \rightarrow D_s \eta$ & $\sim 0.1$ & \\
\hline
$D_{s2}^*(2537) \rightarrow {\rm all}$  &  8 to 11 &  \\[2pt]
$D_{s1}(2502) \rightarrow D^*K$ & 3 to 6 & $< 4.6$ \\[4pt]
$B_2^*(5767) \rightarrow B^*\pi$    & 11 & \\
$B_2^*(5767) \rightarrow B\pi$  & 10 & \\
$B_2^*(5767) \rightarrow B^* \rho$  & 13 to 29 & \\
$B_2^*(5767) \rightarrow B \rho$    & 4 to 13 & \\
\hline
$B_2^*(5767) \rightarrow {\rm all}$ & 38 to 63 & \\[2pt]
$B_1(5755) \rightarrow B^*\pi$  & 14 & \\
$B_1(5755) \rightarrow B^* \rho$    & 11 to 33 & \\
$B_1(5755) \rightarrow B \rho$  & 6 to 8 & \\
\hline
$B_1(5755) \rightarrow {\rm all}$   &  31 to 55 & \\[4pt]

$B_{s2}^*(5846) \rightarrow B^*K$   & 2 to 4 & \\
$B_{s2}^*(5846) \rightarrow BK$ & 1 to 3 & \\
\hline
$B_{s2}^*(5846) \rightarrow {\rm all}$  &  3 to 7 &   \\[2pt]

$B_{s1}(5834) \rightarrow B^*K$ & 1 to 3 & \\[4pt]
$K_2^*(1429) \rightarrow K^*\pi$    & 16 to 22 & 25 \\
$K_2^*(1429) \rightarrow K\pi$  & 35 to 40 & 50 \\
$K_2^*(1429) \rightarrow K \rho$    & 10 to 19 & 9 \\
$K_2^*(1429) \rightarrow K \omega$  & 2 to 4 & 3 \\
\hline
$K_2^*(1429) \rightarrow {\rm all}$ &  63 to 85 \\[2pt]

$K_1(1270) \rightarrow K^*\pi$  & 12 to 21 & 14 \\
$K_1(1270) \rightarrow K \rho$   & 38\tablenote{Input value.} & 38 \\
$K_1(1270) \rightarrow K \omega$ & 9 & 10 \\
\hline
$K_1(1270) \rightarrow {\rm all}$   &  59 to 68
\end{tabular}
    \protect\label{pwavewidths}
\end{table}

\begin{table}

        \caption{Properties of the
        3D$({ \frac{5}{2}})$ heavy-light mesons.}
        \begin{tabular}{lcc}
State & Mass (MeV)  & Width (MeV) \\
\hline
$K_3^*$ & 1770  &       170 to 182       \\
$K_2$ & 1770    &       102 to 126       \\[4pt]
$D_3^*$ & 2830  &       324 to 479       \\
$D_2$ & 2830    &       192 to 279       \\[4pt]
$D_{s3}^*$ & 2880       &       103 to 114       \\
$D_{s2}$ & 2880         &       75 to 97         \\[4pt]
$B_3^*$ & 6148  &       285 to 387       \\
$B_2$ & 6148    &       264 to 372       \\[4pt]
$B_{s3}^*$ & 6198       &       121 to 142       \\
$B_{s2}$ & 6198         &       109 to 133       \\
                \end{tabular}
                \protect\label{shortdwavewidths}

\end{table}


\begin{references}
\bibitem{etacomp} We take the $\eta$ wave function to be
$(u\bar{u}+d\bar{d}-s\bar{s})/\sqrt{3}$.
\end{references}
\end{document}